\documentclass[prd,twocolumn,groupedaddress,showpacs,nofootinbib]{revtex4}
\usepackage{graphicx}
\usepackage{dcolumn}
\usepackage{amssymb}
\usepackage{mathrsfs}
\usepackage{amsmath}
\usepackage{epsfig}
\usepackage[dvips]{color}
\usepackage{hhline}

\begin{document}

\title{Quasi-degenerate dark matter for DAMPE excess and $3.5\,\textrm{keV}$ line}

\author{Pei-Hong Gu}
\email{peihong.gu@sjtu.edu.cn}

\affiliation{School of Physics and Astronomy, Shanghai Jiao Tong
University, 800 Dongchuan Road, Shanghai 200240, China}

\begin{abstract}

We propose a quasi-degenerate dark matter scenario to simultaneously explain the $1.4\,\textrm{TeV}$ peak in the high-energy cosmic-ray electron-positron spectrum reported by the DAMPE collaboration very recently and the $3.5\,\textrm{keV}$ X-ray line observed in galaxies clusters and from the Galactic centre and confirmed by the Chandra and NuSTAR satellites. We consider a dark $SU(2)'\times U(1)'$ gauge symmetry under which the dark matter is a Dirac fermion doublet composed of two $SU(2)'$ doublets with non-trivial $U(1)'$ charges. At one-loop level the two dark fermion components can have a mass split as a result of the dark gauge symmetry breaking. Through the exchange of a mediator scalar doublet the two quasi-degenerate dark fermions can mostly annihilate into the electron-positron pairs at tree level for explaining the $1.4\,\textrm{TeV}$ positron anomaly, meanwhile, the heavy dark fermion can very slowly decay into the light dark fermion with a photon at one-loop level for explaining the $3.5\,\textrm{keV}$ X-ray line. Our dark fermions can be also verified in the direct detection experiments. 

\end{abstract}


\maketitle

\section{Introduction}

The DAMPE satellite has been directly measuring the high-energy cosmic-ray electrons and positrons in the energy range $25\,\textrm{GeV}$ to $4.6\,\textrm{TeV}$ with unprecedentedly high energy resolution and low background. Very recently the DAMPE collaboration released their results where the spectrum seems to have a narrow bump above the background at around $1.4\,\textrm{TeV}$ although its largest part is well fitted by a smoothly broken power-law model \cite{dampe2017}. If a dark matter (DM) particle is expected to account for the DAMPE excess, it should mostly annihilate into the electron-positron pairs. There have been a number of works studying the DAMPE excess \cite{fhsty2017,gh2017,duan2017,yuan2017,fby2017,zzfyf2017,twzz2017,cy2017,gu2017-2,abfz2017,cao2017,dhwy2017,ll2017,hwzz2017,ckk2017,gm2017,niu2017}.

On the other hand, an unknown $3.5\,\textrm{keV}$ line in the spectrum of the cosmic X-ray background has been observed in galaxies clusters \cite{bulbul2014,brlf2014} and has been confirmed by the Chandra and NuSTAR satellites \cite{cappelluti2017,mne2014}. This signal can be understood by the decay of a light DM sterile neutrino into an active neutrino and a photon \cite{pw1982} or the annihilation of a light DM pair into two photons. Alternatively, we can consider the decay of a heavy DM particle into a light DM particle with a photon \cite{gu2013}. In this case, the two DM particles should have a quasi-degenerate mass spectrum. So far a lot of of theoretical models have been proposed for interpreting this $3.5\,\textrm{keV}$ X-ray line\cite{ijt2014,fw2016,hjt2014,jrr2014,hpp2014,ppk2014,cs2014,bo2014,tsuyuki2014,bg2014,ku2014,adg2014,bm2014,dhm2014,ehs2014,modak2015,cy2014,fhr2014,pss2015,acd2015,dpmb2017,abazajian2017,ht2017,rst2017,bkly2017,bcck2017}.

It should be interesting to simultaneously explain the $1.4\,\textrm{TeV}$ positron excess and the $3.5\,\textrm{keV}$ X-ray line in a same DM scenario. For this purpose, we could consider the quasi-degenerate DM scenario where the DM particles mostly annihilate into the electron-positron pairs, meanwhile, the heavy DM particle slowly decays into the light DM particle.

In this paper, we shall introduce a dark $SU(2)'\times U(1)'$ gauge symmetry and then consider two $[SU(2)']$-doublet fermion with non-trivial $U(1)'$ charges to form a dark Dirac fermion doublet. We shall also resort to a mediator scalar doublet for constructing the Yukawa couplings of the dark fermion doublet to the standard model (SM) lepton singlets. By choosing these Yukawa couplings, the dark fermion annihilation can account for the $1.4\,\textrm{Tev}$ positron excess. After the dark symmetry breaking, the two components of the dark fermion doublet can have a mass split at one-loop level. In the presence of two additional mediator scalar singlets, the two components of the mediator scalar doublet can mix with each other and hence can mediate a radiative decay of the heavy dark fermion into the light dark fermion with a photon. The $3.5\,\textrm{keV}$ X-ray line thus can be interpreted. Furthermore, the dark fermions could be tested in the direct detection experiments.

This paper is organized as follows. In Sec. II, we introduce the model. In Sec. III, we study the dark gauge bosons and the mediator scalars. In Sec. IV, we demonstrate the DM properties. Finally, we make a conclusion in Sec. V.

\section{The model}

In addition to the $SU(2)' \times U(1)'$ gauge fields $W'^{a}_{\mu}(a=1,2,3)$ and $B'^{}_{\mu}$, we introduce the following fermions and scalars,
\begin{eqnarray}
\!\!\!\!\begin{array}{c}\chi_{L,R}^{}(1,1,0)(2,+\frac{1}{2})\end{array}\!\!\!\!&=&\!\!\left[\begin{array}{c}\chi_{L1}^{0}\\
[2mm]
\chi_{L2}^{0}
\end{array}\right]\!,~~\delta^{+}_{}(1,1,+1)(1,0),\nonumber\\
[2mm]
\!\!\!\!\begin{array}{c}\xi (1,1,0)(2,+\frac{1}{2})\end{array}\!\!\!\!&=&\!\!\left[\begin{array}{c}\xi_{1}^{0}\\
[2mm]
\xi_{2}^{0}
\end{array}\right]\!,~~\sigma^{0}_{}(1,1,0)(1,+1),\nonumber\\
[2mm]
\!\!\!\!\begin{array}{c}\eta (1,1,+1)(2,+\frac{1}{2})\end{array}\!\!\!\!&=&\!\!\left[\begin{array}{c}\eta_{1}^{+}\\
[2mm]
\eta_{2}^{+}
\end{array}\right]\!,~~\omega^{+}_{}(1,1,+1)(1,+1).\end{eqnarray}
Here and thereafter the first and second brackets following the fields describe the transformations under the SM $SU(3)_c^{}\times SU(2)_L^{} \times U(1)^{}_{Y}$ gauge symmetry and the dark $SU(2)'\times U(1)'$ gauge symmetry, while the numbers or signs in the upper indices of the fields are the electric charges. As we will demonstrate in the following, the neutral fermion doublets $\chi_{L,R}^{}$ serve as the DM particles, the charged scalar doublet $\eta$ mediates the annihilation of the DM into the electron-positron pairs, the other charged scalar singlets $\delta$ and $\omega$ participate in mediating the decay of a heavy DM fermion into a light DM fermion and a monochromatic photon, while the neutral scalar doublet $\xi$ and the neutral scalar singlet $\sigma$ are responsible for completely breaking the dark $SU(2)'\times U(1)'$ gauge symmetry. We hence would like to respectively refer to $\chi_{L,R}^{}$ as the dark fermion doublets, $\eta$, $\delta$ and $\omega$ as the mediator scalar doublet or singlet, $\xi$ and $\sigma$ as the dark Higgs doublet or singlet.

The full Lagrangian should be
\begin{eqnarray}
\label{lar}
\mathcal{L}&=&-\frac{1}{4}W'^{a}_{\mu\nu}W'^{a\mu\nu}-\frac{1}{4}B'^{}_{\mu\nu}B'^{\mu\nu}-\frac{\epsilon}{2}B'^{}_{\mu\nu}B^{\mu\nu}_{}\nonumber\\
[2mm]
&&+ i\bar{\chi}_{L,R}^{}\gamma^\mu_{} D_\mu^{} \chi_{L,R}^{}-m_\chi^{}\left(\bar{\chi}_L^{}\chi_R^{} +\textrm{H.c.}\right)\nonumber\\
[2mm]
&&+(D_\mu^{}\xi)^\dagger_{}D^\mu_{}\xi+(D_\mu^{}\sigma)^\dagger_{}D^\mu_{}\sigma - \mu_\xi^2 \xi^\dagger_{}\xi -\lambda_\xi^{}  \left(\xi^\dagger_{}\xi\right)^2_{}\nonumber\\
[2mm]
&&- \mu_\sigma^2 \sigma^\dagger_{}\sigma -\lambda_\sigma^{}  \left(\sigma^\dagger_{}\sigma\right)^2_{}-\mu_\phi^2\phi^\dagger_{}\phi-\lambda_\phi^{}\left(\phi^\dagger_{}\phi\right)^2_{}\nonumber\\
[2mm]
&&-\lambda_{\xi\sigma}^{}\xi^\dagger_{}\xi\sigma^\dagger_{}\sigma- \lambda_{\xi\phi}^{}\xi^\dagger \xi \phi^\dagger_{}\phi- \lambda_{\sigma\phi}^{}\sigma^\dagger \sigma \phi^\dagger_{}\phi\nonumber\\
[2mm]
&&+(D_\mu^{}\delta)^\dagger_{}D^\mu_{}\delta +(D_\mu^{}\eta)^\dagger_{}D^\mu_{}\eta +(D_\mu^{}\omega)^\dagger_{}D^\mu_{}\omega\nonumber\\
[2mm]
&&-\left(\mu_\delta^2 + \lambda_{\delta\phi}^{}\phi^\dagger_{}\phi+\lambda_{\delta\xi}^{}\xi^\dagger_{}\xi + \lambda_{\delta\sigma}^{}\sigma^\dagger_{}\sigma\right)\delta^\dagger_{}\delta\nonumber\\
[2mm]
&& -\left(\mu_\eta^2 + \lambda_{\eta\phi}^{}\phi^\dagger_{}\phi+\lambda_{\eta\xi}^{}\xi^\dagger_{}\xi + \lambda_{\eta\sigma}^{}\sigma^\dagger_{}\sigma\right)\eta^\dagger_{}\eta\nonumber\\
[2mm]
&&  -\left(\mu_\omega^2 + \lambda_{\omega\phi}^{}\phi^\dagger_{}\phi+\lambda_{\omega\xi}^{}\xi^\dagger_{}\xi + \lambda_{\omega\sigma}^{}\sigma^\dagger_{}\sigma\right)\omega^\dagger_{}\omega\nonumber\\
[2mm]
&&- \kappa^{}_{1}\eta^\dagger_{}\tilde{\xi} \tilde{\xi}^\dagger_{}\eta -\kappa^{}_{2}\eta^\dagger_{}\xi \xi^\dagger_{}\eta - \sqrt{2}\rho_{\delta\eta}^{} \left(\delta\eta^\dagger_{}\xi+\textrm{H.c.}\right) \nonumber\\
[2mm]
&&- \sqrt{2} \rho_{\omega\eta}^{} \left(\omega\eta^\dagger_{}\tilde{\xi}+\textrm{H.c.}\right) - \sqrt{2}\rho_{\omega\delta}^{} \left(\sigma\omega^\dagger_{}\delta+\textrm{H.c.}\right)\nonumber\\
[2mm]
&&-\left(f_{\alpha}^{}\bar{\chi}_L^{}\eta e_{R\alpha}^{}+\textrm{H.c.}\right)\nonumber\\
[2mm]
&&+\textrm{~other~terms~in~the~SM}\,.
\end{eqnarray}
Here $D_\mu^{}$ are the covariant derivatives,
\begin{eqnarray}
\!\!\!\!D_\mu^{}\chi_{L,R}^{}&=&\left(\partial_\mu^{} - i g'^{}_2 \frac{\tau_a^{}}{2} W'^{a}_{\mu}-i \frac{1}{2}g'^{}_{1}B'^{}_\mu  \right) \chi_{L,R}^{}\,,\nonumber\\
[1mm]
\!\!\!\!D_\mu^{}\delta&=&\left(\partial_\mu^{}  - i g' B^{}_\mu \right) \delta\,,\nonumber\\
[1mm]
\!\!\!\!D_\mu^{}\xi&=&\left(\partial_\mu^{} - i g'^{}_2 \frac{\tau_a^{}}{2} W'^{a}_{\mu} -i \frac{1}{2}g'^{}_{1}B'^{}_\mu \right) \xi\,,\nonumber\\
[1mm]
\!\!\!\!D_\mu^{}\sigma&=&\left(\partial_\mu^{} - i g'^{}_{1}B'^{}_\mu \right) \sigma\,,\nonumber\\
[1mm]
\!\!\!\!D_\mu^{}\eta&=&\left(\partial_\mu^{} - i g'^{}_2 \frac{\tau_a^{}}{2} W'^{a}_{\mu}  - i \frac{1}{2}g'^{}_{1}B'^{}_\mu - i g' B^{}_{\mu}\right) \eta \,,\nonumber\\
[1mm]
\!\!\!\!D_\mu^{}\omega&=&\left(\partial_\mu^{}   -  i g'^{}_1 B'^{}_{\mu} -  i g' B^{}_{\mu} \right)\omega \,, \nonumber
 \end{eqnarray}
while $\phi$ and $e_{R\alpha}^{}$ are the SM doublet and lepton singlets, 
\begin{eqnarray}
\!\!\!\!\begin{array}{c}\phi (1,2,+\frac{1}{2})(1,0)\end{array}\!\!\!\!&=&\!\!\left[\begin{array}{c}\phi_{}^{+}\\
[2mm]
\phi_{}^{0}
\end{array}\right]\!,~~e_{R\alpha}^{}(1,1,-1)(1,0)\,.
\end{eqnarray} 
Note we have forbidden the gauge-invariant Yukawa couplings of the mediator scalar singlet $\delta$ to the SM lepton doublets by imposing a $Z_2^{}$ discrete symmetry under which only the dark fermions $\chi_{L,R}^{}$ and the mediator scalars $\eta$, $\omega$ and $\delta$ are odd.

\section{Dark gauge bosons and mediator scalars}

The dark Higgs doublet $\xi$ will be responsible for spontaneously breaking the dark $SU(2)'\times U(1)'$ symmetry down to a dark $U(1)''$ symmetry, which will be eventually broken when the dark Higgs singlet $\sigma$ develops a vacuum expectation value (VEV), i.e.
\begin{eqnarray}
SU(2)'\times U(1)' \stackrel{\langle\xi\rangle}{\longrightarrow}  U(1)'' \stackrel{\langle\omega\rangle}{\longrightarrow} I\,. 
 \end{eqnarray}
 We thus write the dark Higgs scalars $\xi$ and $\sigma$ by 
 \begin{eqnarray}
 \xi = \left[\begin{array}{c} 
 0\\
 [2mm]\frac{1}{\sqrt{2}}\left(v_\xi^{}+h_\xi^{}\right)
 \end{array}\right],~~\sigma =\frac{1}{\sqrt{2}}\left(v_\sigma^{}+h_\sigma^{}\right)\,,
\end{eqnarray} 
with $v_{\xi,\sigma}^{}$ and $h_{\xi,\sigma}^{}$ being the VEVs and the Higgs bosons. The SM Higgs doublet $\phi$ develops its VEV for the electroweak symmetry breaking as usual, i.e.
\begin{eqnarray}
\phi=\left[\begin{array}{c}
0\\
[2mm]
\frac{1}{\sqrt{2}}\left(v_\phi^{}+h_\phi^{}\right)
\end{array}\right]~~\textrm{with}~~ v_\phi^{}=246\,\textrm{GeV}\,.
\end{eqnarray}

As for the charged mediator scalars $\delta^{\pm}_{}$, $\eta^{\pm}_{1,2}$ and $\omega^{\pm}_{}$, their masses should be
\begin{eqnarray}
\mathcal{L} &\supset&-M_\delta^2 \delta^{+}_{}\delta^{-}_{}-M_{\eta_1^{}}^2 \eta_1^{+}\eta_1^{-}-M_{\eta_2^{}}^2 \eta_2^{+} \eta_2^{-}-M_\omega^2\omega^{+}_{}\omega^{-}_{}\nonumber\\
[2mm]
&&-\rho_{\delta\eta}^{}v_\xi^{}\delta^{+}_{}\eta^{-}_{2} -\rho_{\omega\eta}^{}v_\xi^{}\omega^{+}_{}\eta^{-}_{1}-\rho_{\omega\delta}^{}v_\sigma^{}\delta^{+}_{}\omega^{-}_{}\nonumber\\
[2mm]
&&+\textrm{H.c.} ~~\textrm{with}\nonumber\\
[2mm]
&&M_{\delta}^2=\mu_\delta^2 + \frac{1}{2}\left(\lambda_{\delta\phi}^{}v^{2}_\phi+\lambda_{\delta\xi}^{}v_\xi^{2} + \lambda_{\delta\sigma}^{}v_\sigma^{2}\right),\nonumber\\
[2mm]
&&M_{\eta_1^{}}^2=\mu_\eta^2 + \frac{1}{2}\left(\lambda_{\eta\phi}^{}v^{2}_\phi+\lambda_{\eta\xi}^{}v_\xi^{2} +\kappa^{}_{1}v_\xi^2+ \lambda_{\eta\sigma}^{}v_\sigma^{2}\right),\nonumber\\
[2mm]
&&M_{\eta_2^{}}^2=\mu_\eta^2 + \frac{1}{2}\left(\lambda_{\eta\phi}^{}v^{2}_\phi+\lambda_{\eta\xi}^{}v_\xi^{2} +\kappa^{}_{2}v_\xi^2+ \lambda_{\eta\sigma}^{}v_\sigma^{2}\right),\nonumber\\
[2mm]
&&M_\omega^2 =\mu_\omega^2 + \frac{1}{2}\left(\lambda_{\omega\phi}^{}v^{2}_\phi+\lambda_{\omega\xi}^{}v_\xi^{2} + \lambda_{\omega\sigma}^{}v_\sigma^{2}\right).
\end{eqnarray}
Clearly, these charged scalars mix together. If the $\delta^{\pm}_{}$ and $\omega^{\pm}_{}$ scalars are heavy enough, they can be simply integrated out from the theory. In this limiting case, we can obtain a $\eta_{1}^{}-\eta_{2}^{}$ mixing as below,
\begin{eqnarray}
\label{eta}
\mathcal{L} \!&\supset&\!\! - \Delta m_{\eta}^2  \eta_1^{+}\eta_2^{-}+\textrm{H.c.}~~\textrm{with}\nonumber\\
[2mm]
\!&&\!\!\Delta m_{\eta}^2 =\! \left(\frac{\rho_{\delta\eta}^{} v_\xi^{}}{M_\delta^2} \right)\!\!\left( \frac{\rho_{\omega\eta}^{}v_\xi^2 }{M_\omega^2}\right) \rho_{\omega\delta}^{}v_\sigma^{}\ll  \rho_{\omega\delta}^{}v_\sigma^{}\ll M_{\eta_{1,2}^{}}^2\,.
\nonumber\\
\!&&\!\!
\end{eqnarray}
As we will show later the above mixing is essential to realize a decay of a heavy dark fermion into a light dark fermion with a monochromatic photon.

The masses of the $SU(2)'\times U(1)'$ gauge bosons can be given by 
\begin{eqnarray}
\mathcal{L} &\supset& \frac{g'^2_2 v^2_\xi }{4} W'^{+\mu}_{}W'^{-}_{\mu} +\frac{g'^2_2 v^2_\xi}{8\cos^2_{}\!\theta'^{}_W}Z'^{\mu}_{}Z'^{}_{\mu} \nonumber\\
&&+\frac{g'^{2}_{2}\tan^2_{}\!\theta'^{}_W v_\sigma^2}{2}\left(A'^{}_\mu \cos\theta'^{}_W - Z'^{}_\mu \sin\theta'^{}_W\right)^2_{}\,,
\end{eqnarray}
where we have defined 
\begin{eqnarray}
\!\!W'^\pm_{\mu}\!&=& \!\frac{1}{\sqrt{2}}\left(W'^{1}_{\mu}\mp W'^{2}_{\mu}\right)\,,\nonumber\\
[2mm]
\!\!Z'^{}_{\mu}\!&=&\!W'^3_\mu \cos \theta'^{}_W - B'^{}_\mu \sin\theta'^{}_W\,,\nonumber\\
[2mm]
\!\!A'^{}_{\mu}\!&=&\!W'^3_\mu \sin \theta'^{}_W  + B'^{}_\mu \cos\theta'^{}_W\,,~~\tan\theta'^{}_W =\frac{g'^{}_1}{g'^{}_2}\,.
\end{eqnarray}
Clearly, the $W'^{\pm}_\mu$ boson is a mass eigenstate while the $Z'^{}_\mu$ and $A'^{}_\mu$ bosons mix with each other. For simplicity, we take  
\begin{eqnarray}
g'^2_2 v_\xi^2\gg g'^2_1 v_\sigma^2\,,
\end{eqnarray}
and then simplify 
 \begin{eqnarray}
m_{W'}^2 &=& \frac{1}{4} g'^2_2 v^2_\xi\,,~~m_{Z'}^2 \simeq  \frac{1}{4\cos^2_{}\!\theta'^{}_W} g'^2_2 v^2_\xi\,,\nonumber\\
[2mm]
m_{A'}^2 &\simeq&  e'^2_{}v_\sigma^2~~\textrm{with}~~e'= g'^{}_2 \sin\theta'^{}_W\,.
\end{eqnarray}

Because of the $U(1)$ kinetic mixing, the dark photon $A'^{}_\mu$ can couple to the SM fermion pairs at tree level, i.e.
\begin{eqnarray}
\label{kint}
\mathcal{L}\supset \epsilon A'^{}_\mu \left(-\frac{1}{3}\bar{d}\gamma^\mu_{}d +\frac{2}{3}\bar{u}\gamma^\mu_{}u -\bar{e}\gamma^\mu_{}e \right)~~\textrm{for}~~\epsilon \ll 1\,.
\end{eqnarray}  
Through the Yukawa couplings, the dark gauge bosons $X'^{}_\mu=(W'^{\pm}_{},Z'^{}_{\mu\nu},A'^{}_{\mu\nu})$ can couple to the lepton pairs at one-loop level \cite{pw1982}, i.e.
\begin{eqnarray}
\label{darkmagn}
\mathcal{L}&\supset& i \bar{e}_\alpha^{} \left(\frac{W'^{\pm}_{\mu\nu} }{\Lambda_{W'}^{\alpha\beta}}+\frac{Z'^{}_{\mu\nu} }{\Lambda_{Z'}^{\alpha\beta}}+\frac{A'^{}_{\mu\nu} }{\Lambda_{A'}^{\alpha\beta}}\right)\sigma^{\mu\nu}_{} e_\beta^{}~~\textrm{with}\nonumber\\
[2mm]
&&X'^{}_{\mu\nu}=\partial_\mu^{}X'^{}_\nu - \partial_\nu^{} X'^{}_\mu\,,\nonumber\\
[2mm]
&&\frac{1}{\Lambda_{W'}^{\alpha\beta}}
\sim \frac{g'^{}_2 f_{\alpha}^{\ast} f_{\beta}^{}}{16\pi^2_{}}\frac{m_\alpha^{} P_L^{}+m_\beta^{}P_R^{}}{M_{\eta_{1,2}^{}}^2}\,,\nonumber\\
[2mm]
&&\frac{1}{\Lambda_{Z'}^{\alpha\beta}}
\sim \frac{g'^{}_2 f_{\alpha}^{\ast} f_{\beta}^{}}{16\pi^2_{}\cos\theta'^{}_W}\frac{m_\alpha^{} P_L^{}+m_\beta^{}P_R^{}}{M_{\eta_{1,2}^{}}^2}\,,\nonumber\\
[2mm]
&&\frac{1}{\Lambda_{A'}^{\alpha\beta}}
\sim \frac{e' f_{\alpha}^{\ast} f_{\beta}^{}}{16\pi^2_{}}\frac{m_\alpha^{}P_L^{} +m_\beta^{}P_R^{}}{M_{\eta_{1,2}^{}}^2}\,.
\end{eqnarray}

\section{Dark matter properties}

The dark fermion doublets $\chi_{L,R}^{}$ can form the Dirac fermions as below, 
\begin{eqnarray}
\chi_{1}^{}=\chi_{L1}^{}+\chi_{R1}^{}\,,~~\chi_{2}^{}=\chi_{L2}^{}+\chi_{R2}^{}\,,
\end{eqnarray}
which have different $U(1)''$ charges, i.e. $Q''=1$ for $\chi_1^{}$ and $Q''=0$ for $\chi_2^{}$. Clearly, the dark fermion $\chi_1^{}$ other than the dark fermion $\chi_2^{}$ can couple to the dark photon $A'^{}_\mu$. Although the $\chi_1^{}$ and $\chi_2^{}$ components have a same mass $m_\chi^{}$ at tree level, their degeneracy can be broken by the gauge and Yukawa interactions at one-loop level,   
\begin{eqnarray}
\label{split}
\!\!\!\!\!\!\!\!\Delta m_\chi^{}&\equiv& m_{\chi_{1}^{}}^{} - m_{\chi_{2}^{}}^{}  \simeq  \frac{g'^2_2 \sin\theta'^{}_W \tan\theta'^{}_W}{8\pi} m_{W'}^{} \nonumber\\
&&\quad\quad\quad\quad \quad\quad~+ \frac{m_\chi^{}}{16\pi^2_{}}\ln\left(\frac{M_{\eta_2^{}}^2}{M_{\eta_1^{}}^2} \right)\left|f_\alpha^{}\right|^2_{}\nonumber\\
[2mm]
\!\!\!\!\!\!\!\!&&\textrm{for}~M_{\eta_{1,2}^{}}^2\gg m_\chi^{2}\gg m_{W'}^{2}\,,~m_{Z'}^{2}\gg m_{A'}^{2}\,.
\end{eqnarray}
The first and second terms are from the gauge and Yukawa interactions, respectively. The contribution from the gauge interactions is similar to that in the so-called minimal DM scenario \cite{cfs2006}. The mass split can arrive at a small value if the Yukawa and gauge contributions are both small or the two contributions have a large cancellation.

Through the exchange of the mediator scalar doublet $\eta$ and the SM lepton singlets $e_{R\alpha}^{}$, the Yukawa couplings in Eq. (\ref{lar}) can induce a magnetic moment of the dark fermions $\chi_{1,2}^{}$ at one-loop level \cite{pw1982}, i.e.
\begin{eqnarray}
\label{magn1}
\mathcal{L}&\supset& \frac{i}{\Lambda_{1,2}^{}}\bar{\chi}_{1,2}^{}\sigma^{\mu\nu}_{}\chi_{1,2}^{} A_{\mu\nu}^{} \nonumber\\
[2mm]
&&\textrm{with}~~\frac{1}{\Lambda_{1,2}^{}}
\sim \frac{e\left|f_{\alpha}^{} \right|^2_{}}{16\pi^2_{}}\frac{m_\chi^{} }{M_{\eta_{1,2}^{}}^2}\,.
\end{eqnarray}
In Eq. (\ref{eta}), we have shown the $\eta_{1,2}^{}$ components of the mediator scalar doublet $\eta$ can obtain a mixing suppressed by the mediator scalar singlets $\delta$ and $\omega$ after the dark $SU(2)' \times U(1)'$ gauge symmetry is completely broken. In association with the Yukawa couplings in Eq. (\ref{lar}), the mixed $\eta_{1,2}^{}$ components and the SM lepton singlets $e_{R\alpha}^{}$ can mediate a magnetic moment between the two dark fermions $\chi_{1,2}^{}$ at one-loop level \cite{pw1982}, i.e. 
\begin{eqnarray}
\label{magn12}
\mathcal{L}&\supset& \frac{i}{\Lambda_{12}^{}}\bar{\chi}_2^{}\sigma^{\mu\nu}_{}\chi_1^{} A_{\mu\nu}^{} +\textrm{H.c.}\nonumber\\
&&\textrm{with}~~\frac{1}{\Lambda_{12}^{}}
\sim \frac{e\left|f_{\alpha}^{} \right|^2_{}}{16\pi^2_{}}\frac{m_\chi^{} \Delta m_\eta^2 }{M_{\eta_{1}^{}}^2 M_{\eta_{2}^{}}^2}\,.
\end{eqnarray}
In the present model, the mediator scalar singlets $\delta$ and $\omega$ can be very heavy and hence can help us to suppress the above magnetic moment. In consequence, the heavy dark fermion $\chi_1^{}$ can always have a very long lifetime when it decays into the light dark fermion $\chi_2^{}$ with a monochromatic photon.

In order to explain the $3.5\,\textrm{keV}$ X-ray line by the decays,
\begin{eqnarray}
\chi_1^{}\longrightarrow  \chi_2^{} + \gamma~~\textrm{with}~~E_\gamma^{}=\frac{m_{\chi_1^{}}^{2}-m_{\chi_2^{}}^{2}}{2m_{\chi_1^{}}^{}}=3.5\,\textrm{keV}\,,
\end{eqnarray}
the mass split (\ref{split}) should be extremely small, 
\begin{eqnarray}
\!\!\!\!\Delta m_\chi^{}=m_{\chi_1^{}}^{}- m_{\chi_2^{}}^{}\simeq E_\gamma^{}~~\textrm{for}~~m_{\chi_1^{}}^{}\simeq m_{\chi_2^{}}^{}\gg\Delta m_\chi^{} \,.
\end{eqnarray}
For this purpose, we can choose the dark gauge couplings $g'^{}_{2}$, the dark Weinberg angle $\theta'^{}_W$, the dark gauge boson mass $m_{W'}^{}$, and the mediator scalar masses $M_{\eta_{1,2}^{}}^{}$ in Eq. (\ref{split}). For example, we input
\begin{eqnarray}
\label{par1}
&&g'^{}_2=0.0023\,,~~\sin\theta'^{}_W = 0.041\,,~~m_{W'}^{}=10\,\textrm{GeV}\,,\nonumber\\
[2mm]
&&M_{\eta_1^{}}^{2}=M_{\eta_2^{}}^{2}\,.
\end{eqnarray}
In the following demonstration we will simply take
\begin{eqnarray}
M_{\eta_1^{}}^{2}=M_{\eta_2^{}}^{2}=M_\eta^2\,.
\end{eqnarray}

We now check if the dark fermions $\chi_{1,2}$ can really serve as the DM. From Eq. (\ref{lar}), we can see through the $t$-channel exchange of the mediator scalar doublet $\eta$, the dark fermions $\chi_{1,2}^{}$ can annihilate into the SM lepton singlets $e_{R\alpha}^{}$. The effective cross section is given by 
\begin{eqnarray}
\langle\sigma\left(\chi\bar{\chi}\rightarrow e_{R\alpha}^{}\bar{e}_{R\beta}^{}\right)v_{\textrm{rel}}^{}\rangle & = &\frac{\left|f_{\alpha}^{}f_{\beta}^\ast\right|^2_{}}{32\pi}\frac{m_\chi^2}{M_\eta^4}\nonumber\\
&& \textrm{for}~~m_\chi^{2}\ll M_\eta^{2}\,.
\end{eqnarray}
The positron anomaly reported by the DAMPE implies the dark fermions should mostly annihilate into the electron-positron pairs. In consequence, the Yukawa couplings $f_{e,\mu,\tau}^{}$ should have the following hierarchy, 
\begin{eqnarray}
\left|f_{e}^{} \right|^2_{} \gg \left| f_{\mu,\tau}^{} \right|^2_{}\,.
\end{eqnarray}
We then can take 
\begin{eqnarray}
\label{annep}
\!\!\!\!\!\!\!\!\!\!\!\!\!\!\!\!\langle\sigma\left(\chi\bar{\chi}\rightarrow e_{}^{+}e_{}^{-}\right)v_{\textrm{rel}}^{}\rangle & = &1\,\textrm{pb}\left(\frac{m_\chi^{}}{1.4\,\textrm{TeV}}\right)^2_{}\nonumber\\
[2mm]
\!\!\!\!\!\!\!\!&&\times \left(\frac{\left|f_e^{}\right|}{\sqrt{4\pi}}\right)^2_{}\left(\frac{5.9\,\textrm{TeV}}{M_\eta^{}}\right)^4_{}.
\end{eqnarray}
In addition, the dark $SU(2)'\times U(1)'$ gauge interactions can also contribute to the DM annihilation. However, their contribution can be safely ignored by taking the dark gauge couplings $g'^{}_{1,2}$ small enough and/or the dark gauge bosons $(W'^{\pm}_{\mu}, Z'^{}_{\mu},A'^{}_{\mu})$ heavy enough. Therefore, the annihilating cross section (\ref{annep}) can account for the DM relic density \cite{patrignani2016} as well as the DAMPE excess \cite{dampe2017}.

The heavy dark fermion $\chi_{1}^{}$ with a $U(1)''$ charge $Q''=1$ can couple to the dark photon $A'^{}_\mu$ and hence can scatter off the protons in nuclei at tree level through the $t$-channel exchange of the dark photon. Meanwhile, a virtual ordinary photon can mediate the scattering of the dark fermions $\chi_{1,2}^{}$ off the protons at one-loop level. 
The effective operators at quark level for the scattering should be \cite{bth2011}
\begin{eqnarray}
\mathcal{L}&\supset& a_X^{} Q_q^{}\bar{q}\gamma^\mu_{}q \bar{\chi}_1^{}\gamma_\mu^{} \chi_1^{}+a_{\gamma}^{}Q_q^{}\bar{q}\gamma^\mu_{}q \bar{\chi}_{1,2}^{}\gamma_\mu^{} P_L^{}\chi_{1,2}^{} ~~\textrm{with}\nonumber\\
[2mm]
&&a_X^{}= \frac{ \epsilon e e'}{m_X^2}  \,,~~a_{\gamma}^{}= \frac{e^2_{}\left|f_\alpha^{}\right|^2_{}}{16\pi^2_{} M_\eta^2} \left[\frac{1}{2} +\frac{1}{3}\ln\left(\frac{m_{\alpha}^2}{M_{\eta}^2}\right)\right]  \,,\nonumber\\
&&
\end{eqnarray}
with $Q_q^{}$ being the electric charge of the quark $q$. The scattering cross section can be computed by
\begin{eqnarray}
&&\sigma_{\chi_1^{} p\rightarrow \chi_1^{} p}^{}= \frac{\left(a_X^{}+\frac{1}{2}a_\gamma^{}\right)^2  \mu_r^2}{2\pi} \,,~~\sigma_{\chi_2^{} p\rightarrow \chi_2^{} p}^{}= \frac{a_\gamma^{2} \mu_r^2}{8\pi} \,,\nonumber\\
[2mm]
&&\textrm{with}~~\mu_r^{}=\frac{m_\chi^{}m_p^{}}{m_\chi^{}+m_p^{}}\,.
\end{eqnarray}
By inputting,
\begin{eqnarray}
\label{par2}
\!\!\!\!\!\!\!\!\!\!\!\!&&\left|f_e^{}\right|=\sqrt{4\pi}\gg \left|f_{\mu,\tau}^{}\right|\,,~~M_\eta^{}=5.9\,\textrm{TeV}\,,\nonumber\\
[2mm]
\!\!\!\!\!\!\!\!\!\!\!\!&&\epsilon=10^{-5}_{}\,,~~e'=0.0023\times 0.041\,,~~m_X^{}=250\,\textrm{MeV}\,,~~\end{eqnarray}
we can obtain
\begin{eqnarray}
&&a_X^{}= 4.5\times 10^{-3}_{}\,\textrm{TeV}^{-2}_{}\,,~~
a_\gamma^{}=-1.9\times 10^{-3}_{}\,\textrm{TeV}^{-2}_{},\nonumber\\
[2mm]
&&\sigma_{\chi_1^{} p\rightarrow \chi_1^{} p}^{}=4.2\times 10^{-46}_{}\,\textrm{cm}^2_{}\,,\nonumber\\
[2mm]
&&\sigma_{\chi_2^{} p\rightarrow \chi_2^{} p}^{}=3.5\times 10^{-47}_{}\,\textrm{cm}^2_{}\,,
\end{eqnarray}
which has a potential to be verified in the DM direct detection experiments \cite{lcj2017}. 

It is also easy to check that through their couplings with the SM fermions (\ref{kint}) and (\ref{darkmagn}), the dark gauge bosons can decay before the BBN epoch. Actually, for the previous parameter choice in Eqs. (\ref{par1}) and (\ref{par2}), we can estimate 
\begin{eqnarray}
\Gamma_{W'}^{}&\sim& \frac{m_{W'}^3}{4\pi \left(\Lambda_{W'}^{\tau\tau}\right)^2_{}}\sim 5.8\times 10^{-24}_{}\,\textrm{GeV}\left(\frac{\left|f_{\tau\tau}^{}\right|}{0.6}\right)^4_{}\,,\nonumber\\
[2mm]
 \Gamma_{Z'}^{}&\sim &\frac{m_{Z'}^3}{4\pi \left(\Lambda_{Z'}^{\tau\tau}\right)^2_{}}\sim  5.8\times 10^{-24}_{}\,\textrm{GeV}\left(\frac{\left|f_{\tau\tau}^{}\right|}{0.6}\right)^4_{}\,,\nonumber\\
 [2mm]
\Gamma_{A'}^{}&\sim& \frac{\epsilon^2_{}m_{A'}^{}}{4\pi }\sim 2\times 10^{-12}_{}\,\textrm{GeV}\,. \end{eqnarray}

\section{Conclusion}

In this paper we have demonstrated a quasi-degenerate DM scenario to simultaneously explain the $1.4\,\textrm{TeV}$ DAMPE peak and the $3.5\,\textrm{keV}$ X-ray line. Specifically we introduce a dark $SU(2)'\times U(1)'$ gauge symmetry under which the DM particle is a Dirac fermion doublet composed of two dark doublets. Through the $t$-channel exchange of a mediator scalar doublet, the two quasi-degenerate dark fermions can mostly annihilate into the electron-positron pairs at tree level for explaining the $1.4\,\textrm{TeV}$ DAMPE peak. At one-loop level the spontaneous breaking of the dark gauge symmetry can induce a small mass split between the two dark fermion components. The dark symmetry breaking can also lead to a mixing between the two components of the mediator scalar doublet. The mixing components of the mediator scalar doublet can mediate a radiative decay of the heavy dark fermion into the light fermion with a photon and hence we can understand the $3.5\,\textrm{keV}$ X-ray line. The dark fermions can be also verified in the direct detection experiments.

\textbf{Acknowledgement}: This work was supported by the National Natural Science Foundation of China under Grant No. 11675100 and the Recruitment Program for Young Professionals under Grant No. 15Z127060004.


\begin{thebibliography}{99}



\bibitem{dampe2017}
G. Ambrosi  {\it et al.}, (DAMPE Collaboration), doi:10.1038/nature24475.





\bibitem{tw1990}
e.g. M. Turner and F. Wilczek, Phys. Rev. D \textbf{42}, 1001 (1990); G. Bertone, D. Hooper,  and J. Silk, Phys. Rep. \textbf{405}, 279 (2005); J.L. Feng,  Annu. Rev. Astron. Astrophys. \textbf{48}, 495 (2010).




\bibitem{fhsty2017}
Y.Z. Fan, W.C. Huang, M. Spinrath, Y.L.S. Tsai, and Q. Yuan, arXiv:1711.10995 [hep-ph].

\bibitem{gh2017}
P.H. Gu and X.G. He, arXiv:1711.11000 [hep-ph].

\bibitem{duan2017}
G.H. Duan {\it et al.}, arXiv:1711.11012 [hep-ph].

\bibitem{yuan2017}
Q. Yuan {\it et al.}, arXiv:1711.10989 [astro-ph.HE].


\bibitem{fby2017}
K. Fang, X.J. Bi, and P.F. Yin, arXiv:1711.10996 [astro-ph.HE].






\bibitem{zzfyf2017}
L. Zu, C. Zhang, L. Feng, Q. Yuan, and Y.Z. Fan, arXiv:1711.11052 [hep-ph].

\bibitem{twzz2017}
Y.L. Tang, L. Wu, M. Zhang, and R. Zheng, arXiv:1711.11058 [hep-ph].

\bibitem{cy2017}
W. Chao and Q. Yuan, arXiv:1711.11182 [hep-ph].


\bibitem{gu2017-2}
P.H. Gu, arXiv:1711.11333 [hep-ph].


\bibitem{abfz2017}
P. Athron, C. Balazs, A. Fowlie, and Y. Zhang, arXiv:1711.11376 [hep-ph].


\bibitem{cao2017}
J. Cao  {\it et al.}, arXiv:1711.11452 [hep-ph].


\bibitem{dhwy2017}
G.H. Duan, X.G. He, L. Wu, and J.M. Yang, arXiv:1711.11563 [hep-ph].


\bibitem{ll2017}

X. Liu and Z. Liu, arXiv: 1711.11579 [hep-ph].


\bibitem{hwzz2017}
X.J. Huang, Y.L. Wu, W.H. Zhang, and Y.F. Zhou, arXiv:1712.00005 [astro-ph.HE]


\bibitem{ckk2017}
I. Cholis, T. Karwal, and M. Kamionkowski, arXiv:1712.00011[astro-ph.HE].


\bibitem{gm2017}
Y. Gao and Y.Z. Ma, arXiv:1712.00370 [astro-ph.HE].



\bibitem{niu2017}
J.S. Niu {\it et al.}, arXiv:1712.00372 [astro-ph.HE].





\bibitem{bulbul2014}
E. Bulbul {\it et al.}, Astrophys. J. \textbf{789}, 13 (2014).

\bibitem{brlf2014}
A. Boyarsky, O. Ruchayskiy, D. Iakubovskyi, and J. Franse, Phys. Rev. Lett. \textbf{113}, 251301 (2014).


\bibitem{cappelluti2017}
N. Cappelluti {\it et al.}, arXiv:1701.07932 [astro-ph.CO].

\bibitem{mne2014}
D. Malyshev, A. Neronov, and D. Eckert, Phys. Rev. D \textbf{90}, 103506 (2014).

\bibitem{acb2014}
M. E. Anderson, E. Churazov and J. N. Bregman, Mon. Not. Roy. Astron. Soc. \textbf{452}, 3905 (2015).

\bibitem{pw1982}
P.B. Pal and L. Wolfenstein, Phys. Rev. D \textbf{25}, 766 (1982).


\bibitem{gu2013}
P.H. Gu, Phys. Dark Univ. \textbf{2}, 35 (2013).






\bibitem{ijt2014}
H. Ishida, K. S. Jeong, and F. Takahashi, Phys. Lett. B \textbf{732}, 196 (2014).


\bibitem{fw2016}
D.P. Finkbeiner and N. Weiner, Phys. Rev. D \textbf{94}, 083002 (2016).

\bibitem{hjt2014}
T. Higaki, K. S. Jeong, and F. Takahashi, Phys. Lett. B \textbf{733}, 25 (2014).

\bibitem{jrr2014}
J. Jaeckel, J. Redondo, and A. Ringwald, Phys. Rev. D \textbf{89}, 103511 (2014).


\bibitem{hpp2014}
H.M. Lee, S.C. Park, and W.I. Park, Eur. Phys. J. C \textbf{74}, 3062 (2014).


\bibitem{ppk2014}
J.C. Park, S.C. Park and K. Kong, Phys. Lett. B \textbf{733}, 217 (2014).

\bibitem{cs2014}
K.Y. Choi and O. Seto, Phys. Lett. B \textbf{735}, 92 (2014).

\bibitem{bo2014}
S. Baek and H. Okada, arXiv:1403.1710 [hep-ph].

\bibitem{tsuyuki2014}
T. Tsuyuki, Phys. Rev. D \textbf{90}, 013007 (2014).


\bibitem{bg2014}
F. Bezrukov and D. Gorbunov, Phys. Lett. B \textbf{736}, 494 (2014).

\bibitem{ku2014}
C. Kolda and J. Unwin, Phys. Rev. D \textbf{90}, 023535 (2014). 


\bibitem{adg2014}
R. Allahverdi, B. Dutta, and Y. Gao, Phys. Rev. D \textbf{89}, 127305 (2014).

\bibitem{bm2014}
K.S. Babu and R.N. Mohapatra, Phys. Rev. D \textbf{89}, 115011 (2014).












\bibitem{dhm2014}
E. Dudas, L. Heurtier, and Y. Mambrini, Phys. Rev. D \textbf{90}, 035002 (2014).

\bibitem{ehs2014}
C. El Aisati, T. Hambye, and T. Scarn\`{a}, JHEP \textrm{1408}, 133 (2014). 

\bibitem{modak2015}
K.P. Modak, JHEP \textbf{1503}, 064 (2015). 


\bibitem{cy2014}
C.W. Chiang and T. Yamada, JHEP \textbf{1409}, 006 (2014).

\bibitem{fhr2014}
A. Falkowski, Y. Hochberg, and J.T. Ruderman, JHEP \textbf{1411}, 140 (2014). 


\bibitem{pss2015}
S. Patra, N. Sahoo, and N. Sahu, Phys. Rev. D \textbf{91}, 115013 (2015).

\bibitem{acd2015}
G. Arcadi, L. Covi, and F. Dradi, JCAP \textbf{1507}, 023 (2015).

\bibitem{dpmb2017}
A. Dutta Banik, M. Pandey, D. Majumdar, and A. Biswas, Eur. Phys. J. C \textbf{77}, 657 (2017). 

\bibitem{abazajian2017}
K.N. Abazajian, arXiv:1705.01837 [hep-ph].


\bibitem{ht2017}
J. Heeck and D. Teresi, arXiv:1706.09909 [hep-ph].

\bibitem{rst2017}
L. Roszkowski, E.M. Sessolo, and S. Trojanowski, arXiv:1707.06277 [hep-ph].

\bibitem{bkly2017}
K. J. Bae, A. Kamada, S. P. Liew, and K. Yanagi, arXiv:1707.06418 [hep-ph].


\bibitem{bcck2017}
A. Biswas, S. Choubey, L. Covi, and S. Khan, arXiv:1711.00553 [hep-ph].




\bibitem{cfs2006}
M. Cirelli, N. Fornengo, and A. Strumia, Nucl. Phys. B \textbf{753}, 178 (2006).

 

 
\bibitem{patrignani2016}
C. Patrignani {\it et al.}, (Particle Data Group Collaboration), Chin. Phys. C \textbf{40}, 100001 (2016).



\bibitem{bth2011}
B. Ren, K. Tsumura, and X.G. He, Phys. Rev. D \textbf{84}, 073004 (2011).




\bibitem{lcj2017}
J. Liu, X. Chen, and X. Ji, Nature Physics \textbf{13}, 212 (2017).

\end{thebibliography}
\end{document}